\documentclass[aps,prd,onecolumn,groupedaddress,showpacs,nofootinbib,amssymb]{revtex4-2}
\usepackage[dvips]{graphicx}
\usepackage{amssymb}
\usepackage{amsmath}
\usepackage{graphicx,,color}
\usepackage{amsfonts}
\usepackage{bm}
\usepackage{cancel}
\usepackage{comment}

\newcommand\be{\begin{equation}}
\newcommand\ee{\end{equation}}

\newcommand\e{\mathrm{e}}

\allowdisplaybreaks[4]

\begin{document}

\tolerance=5000
\title{Kinetic Axion Dark Matter in String Corrected $f(R)$ Gravity}
\author{V.K. Oikonomou$^{1,2}$}
\email{Corresponding author:
voikonomou@gapps.auth.gr;v.k.oikonomou1979@gmail.com,voikonomou@auth.gr}
\author{F.P. Fronimos$^{1}$}\,
\email{fotisfronimos@gmail.com}
\author{Pyotr Tsyba$^{2}$}\,
\email{pyotrtsyba@gmail.com}
\author{Olga Razina$^{2}$}\,
\email{olvikraz@mail.ru} \affiliation{$^{1)}$Department of
Physics, Aristotle University of Thessaloniki, Thessaloniki 54124,
Greece} \affiliation{$^{2)}$L.N. Gumilyov Eurasian National
University - Astana, 010008, Kazakhstan}

\tolerance=5000

\begin{abstract}
Under the main assumption that the axion scalar field mainly
composes the dark matter in the Universe, in this paper we shall
extend the formalism of kinetic axion $R^2$ gravity to include
Gauss-Bonnet terms non-minimally coupled to the axion field. As we
demonstrate, this non-trivial Gauss-Bonnet term has dramatic
effects on the inflationary phenomenology and on the kinetic axion
scenario. Specifically, in the context of our formalism, the
kinetic axion ceases to be kinetically dominated at the end of the
inflationary era, since the condition $\dot{\phi}\simeq 0$
naturally emerges in the theory. Thus, unlike the case of kinetic
axion $R^2$ gravity, the Gauss-Bonnet corrected kinetic axion
$R^2$ gravity leads to an inflationary era which is not further
extended and the reheating era commences right after the
inflationary era, driven by the $R^2$ fluctuations.
\end{abstract}


\maketitle

\section{Introduction}

Dark matter is still mysterious to date, since no evidence of this
elusive dark component of our Universe has been found to date. The
particle nature of dark matter is very well motivated, since
evidence of dark matter halos around spiral galaxies, and cosmic
events like the bullet cluster point to a particle nature of dark
matter. In the past, many candidate particle were proposed to
describe particle dark matter
\cite{Bertone:2004pz,Bergstrom:2000pn,Mambrini:2015sia,Profumo:2013yn,Hooper:2007qk,Oikonomou:2006mh},
to date no evidence for a dark matter particle has ever been
found. This is probably because earlier studies and experiments
focused to Weakly Interacting Massive Particles (WIMPs), with
masses ranging from 100$\,$MeV to hundreds GeV. Thus currently the
interest of cosmologists is turned to light dark matter particle
candidates. The prominent candidate for a large amount of
phenomenological reasons is the axion, see Refs.
\cite{Preskill:1982cy,Abbott:1982af,Dine:1982ah,Marsh:2015xka,Chadha-Day:2021szb,Choi:2020rgn,DiLuzio:2020wdo,Sikivie:2006ni,Raffelt:2006cw,Linde:1991km,Co:2019jts,Co:2020dya,Barman:2021rdr,Marsh:2017yvc,Odintsov:2019mlf,Nojiri:2019nar,Nojiri:2019riz,Odintsov:2019evb,Cicoli:2019ulk,Fukunaga:2019unq,Caputo:2019joi,maxim,Chang:2018rso,Irastorza:2018dyq,Anastassopoulos:2017ftl,Sikivie:2014lha,Sikivie:2010bq,Sikivie:2009qn,Caputo:2019tms,Masaki:2019ggg,Soda:2017sce,Soda:2017dsu,Aoki:2017ehb,Masaki:2017aea,Aoki:2016kwl,Obata:2016xcr,Aoki:2016mtn,Ikeda:2019fvj,Arvanitaki:2019rax,Arvanitaki:2016qwi,Arvanitaki:2014wva,Arvanitaki:2014dfa,Sen:2018cjt,Cardoso:2018tly,Rosa:2017ury,Yoshino:2013ofa,Machado:2019xuc,Korochkin:2019qpe,Chou:2019enw,Chang:2019tvx,Crisosto:2019fcj,Choi:2019jwx,Kavic:2019cgk,Blas:2019qqp,Guerra:2019srj,Tenkanen:2019xzn,Huang:2019rmc,Croon:2019iuh,Day:2019bbh,Odintsov:2020iui,Nojiri:2020pqr,Odintsov:2020nwm,Oikonomou:2020qah,Oikonomou:2022ela,Oikonomou:2022tux,Mazde:2022sdx,Chen:2022nbb,Caloni:2022uya,Roy:2021uye,DiLuzio:2021qct,Choi:2021aze,DiLuzio:2021pxd,Bauer:2020jbp,Ramberg:2020oct,DiLuzio:2020jjp,Visinelli:2018utg,Visinelli:2018wza,DiLuzio:2016sbl},
for an important stream of articles and reviews, and also
\cite{Semertzidis:2021rxs,Chadha-Day:2021szb} for some recent
updated reviews. Some interesting simulations can also be found in
\cite{Buschmann:2021sdq} which predict a mass for the axion of the
$\mu$eV order and in \cite{BREAD:2021tpx} an experimental proposal
was introduced. Furthermore, an interesting explanation of the
recent Gamma ray bursts observations
\cite{Hoof:2022xbe,Li:2022pqa} can be provided by axion-like
particles with masses of the order $m_a\sim
\mathcal{O}(10^{-10})\,$eV. Among axion models, the most important
and phenomenologically appealing are the ones that have their
primordial Peccei-Quinn $U(1)$ symmetry broken during inflation,
unlike in the QCD axion. These belong to the class of misalignment
axion models, with two characteristic candidates, the canonical
misalignment model \cite{Marsh:2015xka} and the kinetic axion
model \cite{Co:2019jts,Co:2020dya,Barman:2021rdr}. In both models,
the axion is misaligned from the minimum of its potential, however
in the canonical misalignment has no initial kinetic energy, while
in the kinetic misalignment it possesses a large kinetic energy.
Hence, this excess in the kinetic energy delays the axion
oscillations which start when the axion reaches its minimum and
when the axion mass $m_a$ becomes of the same order as the Hubble
rate, that is $m_a\sim H$.

In a previous work \cite{Oikonomou:2022tux} we studied the effects
of the kinetic axion on the $R^2$ inflation theory. As we showed,
although the axion effects are insignificant during the
inflationary era, the kinetic axion dominates the early
post-inflationary era, dominating over the $\langle R^2\rangle $
fluctuations, which effectively destabilize the quasi-de Sitter
vacuum $R^2$ attractor. Thus post-inflationary the Universe
experiences a short stiff era controlled by the axion, before the
latter settles in the minimum of its potential, starting its
oscillations and having an energy density redshifting as dark
matter. In this article, we aim to study the effects of string
corrections in the kinetic axion $R^2$ Lagrangian, in the form of
Einstein-Gauss-Bonnet corrections. The motivation to have a scalar
field and considering modified gravity corrections of the form of
higher curvature invariants comes from the fact that the standard
four dimensional vacuum configuration scalar field Lagrangian,
\begin{equation}\label{generalscalarfieldaction}
\mathcal{S}_{\varphi}=\int
d^4x\sqrt{-g}\left(\frac{1}{2}Z(\varphi)g^{\mu
\nu}\partial_{\mu}\varphi
\partial_{\nu}\varphi+\mathcal{V}(\varphi)+h(\varphi)\mathcal{R}
\right)\, ,
\end{equation}
in which the scalar field must be either conformally or minimally
coupled, receives the following one loop quantum corrections
\cite{Codello:2015mba,Oikonomou:2022bqb}
\begin{align}\label{quantumaction}
&\mathcal{S}_{eff}=\int
d^4x\sqrt{-g}\Big{(}\Lambda_1+\Lambda_2
\mathcal{R}+\Lambda_3\mathcal{R}^2+\Lambda_4 \mathcal{R}_{\mu
\nu}\mathcal{R}^{\mu \nu}+\Lambda_5 \mathcal{R}_{\mu \nu \alpha
\beta}\mathcal{R}^{\mu \nu \alpha \beta}+\Lambda_6 \square
\mathcal{R}\\ \notag &
+\Lambda_7\mathcal{R}\square\mathcal{R}+\Lambda_8 \mathcal{R}_{\mu
\nu}\square \mathcal{R}^{\mu
\nu}+\Lambda_9\mathcal{R}^3+\mathcal{O}(\partial^8)+...\Big{)}\, ,
\end{align}
with the parameters $\Lambda_i$, $i=1,2,...,6$ being some
dimensionful constants, see also the interesting Ref.
\cite{Lambiase:2022ucu} for extra Chern-Simons corrections on the
quantum action. The above corrections contain fourth order
derivatives and the action is compatible with diffeomorphism
invariance. Apparently, in our case the axion scalar field during
inflation is described by $Z(\varphi)=-1$ and $h(\varphi)=1$ in
the action (\ref{generalscalarfieldaction}). Having modified
gravity in its various forms
\cite{reviews1,reviews2,reviews3,reviews4,reviews5} driving
inflation and possibly the dark energy era, a viable framework is
offered in which inflation and the dark energy era may be
described by the same theory
\cite{Nojiri:2003ft,Nojiri:2007as,Nojiri:2007cq,Cognola:2007zu,Nojiri:2006gh,Appleby:2007vb,Elizalde:2010ts,Odintsov:2020nwm,Sa:2020fvn}.
Before we proceed, it should be stated that throughout this paper,
a homogeneous and isotropic background shall be considered with
the line element being,
\begin{equation}
\centering
\label{metric}
ds^2=-dt^2+a^2(t)\delta_{ij}dx^idx^j\, ,
\end{equation}
where $a(t)$ stands for the scale factor. In turn, the axion field
is assumed to be homogeneous as well therefore hereafter,
$\phi=\phi(t)$. This simplifies a lot the subsequent calculations
and it is a well motivated assumption.

\section{kinetic axion $f(R)$ gravity with string corrective terms}

In the literature, the axion dynamics have been investigated for a
plethora of models, since it is a prominent candidate for
non-thermal dark matter. In this paper, based on this assumption,
we shall investigate the inflationary phenomenology of the kinetic
axion model in the presence of higher order curvature invariants.
In particular, we shall make use of an $f(R)$ model which is known
for producing viable inflationary phenomenology, while furthermore
a non-minimal coupling between the axion and the Gauss-Bonnet
density shall be considered, in order to explicitly have a
coupling between the scalar field and curvature. Hence, in order
to study the primordial era of our Universe, the following
gravitational action is proposed,
\begin{equation}
\centering
\label{action}
\mathcal{S}=\int d^4x\sqrt{-g}\bigg(\frac{f(R)}{2\kappa^2}-\frac{1}{2}g^{\mu\nu}\nabla_\mu\phi\nabla_\nu\phi-V(\phi)-\xi(\phi)\mathcal{G}\bigg)\, ,
\end{equation}
where $R$ is the Ricci scalar, $\kappa=\frac{1}{M_P}$ with
$M_P=1/\sqrt{8\pi G}$ being the reduced Planck mass,
$\frac{1}{2}g^{\mu\nu}\nabla_\mu\phi\nabla_\nu\phi$ and $V(\phi)$
stand for the kinetic term and the canonical scalar field
potential, while $\xi(\phi)$ is the arbitrary for the time being
scalar coupling function of the axion to the Gauss-Bonnet
invariant, which shall be specified subsequently and
$\mathcal{G}=R^{\mu\nu\rho\sigma}R_{\mu\nu\rho\sigma}-4R^{\mu\nu}R_{\mu\nu}+R^2$
is the Gauss-Bonnet invariant with $R_{\mu\nu}$ and
$R_{\mu\nu\rho\sigma}$ being the Ricci and Riemann tensor
respectively. In this model, there are two tensor degrees of
freedom, and an additional scalar mode coming from the $f(R)$
gravity. Here, it should be stated that in order to have a direct
impact of the Gauss-Bonnet density in the overall phenomenology, a
non-trivial coupling is required due to the fact that the model is
studied in $D=4$, hence the reason why the arbitrary function
$\xi(\phi)$ is introduced. This inclusion has a major impact on
the overall phenomenology as we shall showcase explicitly. It
should also be stated that recently, an additional approach has
been considered where by working initially in $D$ dimensions and
introducing the Gauss-Bonnet density in the gravitational action
without a non-trivial coupling but with a constant factor scaling
as $\frac{\alpha}{D-4}$, it is possible to keep such contribution
in the limit $D\to4$ as it was shown in Ref. \cite{Ai:2020peo},
however this approach, although worthy of being mentioned, shall
not be considered here. This is because the non-minimal coupling
between the scalar field and the Gauss-Bonnet density in this case
affects the continuity equation of the scalar field compared to
the case of \cite{Oikonomou:2022tux}, and thus a completely new
territory is explored. Before we proceed with the equations of
motion and the inflationary phenomenology, a quick statement on
the Gauss-Bonnet term and its impact on tensor perturbations
should be made.

\begin{figure}[h!]
\centering
\includegraphics[width=20pc]{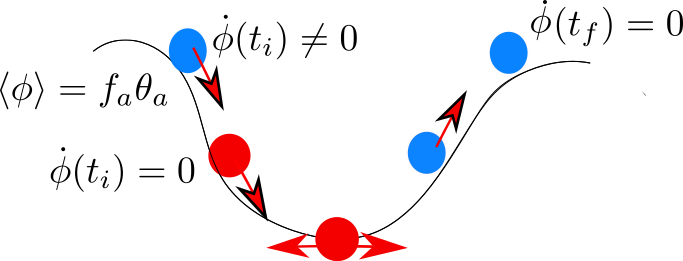}
\caption{The canonical and kinetic misalignment models.}
\label{pictorialrepresentationaxion}
\end{figure}

By including the Gauss-Bonnet density in Eq. (\ref{action}), the
continuity equation of the scalar field, and in consequence the
scalar perturbations, are not the only objects that are affected.
This inclusion is known for being a subclass of Horndeski's theory
and thus the propagation velocity of tensor perturbations is also
influenced by such term. According to Ref. \cite{Hwang:2005hb},
one can show that the deviation of the propagation speed of tensor
perturbations from the speed of light is quantified by the
following expression,
\begin{equation}
\centering
\label{cT}
c_\mathcal{T}^2=1-\frac{Q_f}{Q_t}\, ,
\end{equation}
where $Q_f=8(\ddot\xi-H\dot\xi)$ and
$Q_t=\frac{1}{\kappa^2}\frac{df}{dR}-8\dot\xi H$ are auxiliary
parameters. Recently, the GW170817 event indicated explicitly that
gravitational waves propagate throughout spacetime with the
velocity of light, therefore theories which predict a different
result have been excluded. In this case however, one can postulate
that the arbitrary Gauss-Bonnet scalar coupling function is not so
arbitrary but it needs to satisfy the differential equation,
\begin{equation}
\centering
\label{difeq}
\ddot\xi=H\dot\xi\, ,
\end{equation}
which in turn implies that the auxiliary parameter $Q_f$ in Eq.
(\ref{cT}) vanishes identically. As a result, the Gauss-Bonnet
models that respect this constraint can actually be in agreement
with the latest observations. At this point, one may argue that
the constraint is not necessarily needed because the behavior of
$\phi$, and in consequence $\xi(\phi)$ differs between early and
late-time. It could be the case that while primordially the scalar
field has a dynamical evolution that in consequence predicts
$c_\mathcal{T}^2<1$, the scalar field freezes at late times,
resulting in the conditions $\dot\xi=0=\ddot\xi$ and thus being in
agreement with the GW170817 event in the late-time era only. But
it hard to see how this statement could be realized in the
standard cosmological evolution of the Universe post-inflationary.
There is no fundamental reasoning emanating from particle physics
that could allow the graviton to be massive initially and only at
late times it becomes massless. Thus in the present article we
shall avoid this argument and focus solely on the inflationary
phenomenology of the constrained Gauss-Bonnet model. The
Gauss-Bonnet model is a string inspired model and serves as a
low-energy effective theory, so during and after inflation, the
gravitons should be massless. Hence our massless graviton approach
is well justified. Thus the constraint on the velocity of tensor
perturbations shall be implemented in order to predict massless
gravitons throughout the evolution of the Universe. Therefore, the
constraint (\ref{difeq}) will have a major impact on the
continuity equation and in turn, it will influence the evolution
of the scalar field. In Ref. \cite{Odintsov:2020sqy}, it was
proved that the constraint decreases the overall degrees of
freedom and one can treat the continuity equation of the scalar
field as a differential equation from which a scalar function,
either the potential or the Gauss-Bonnet scalar coupling function,
are specified. Here, since we know the behaviour of the kinetic
axion model, the usual approach shall be considered where the
derivative of the scalar field is specified from the continuity
equation once the Gauss-Bonnet scalar coupling function is
designated. This is possible due to the fact that Eq.
(\ref{difeq}) can be rewritten by making use of the chain rule
$\dot\xi=\xi'\dot\phi$ as,
\begin{equation}
\centering
\label{difeq2}
\ddot\phi=H\dot\phi\bigg[1-\frac{\xi''}{H\xi'}\dot\phi\bigg]\, ,
\end{equation}
where for simplicity, differentiation with respect to the scalar
field is denoted with the ``prime''. This equation is of paramount
importance, since not only does it facilitate the derivation of
$\dot\phi$ from the continuity equation, but it is also connected
to one of the slow-roll indices that we are interested in, as we
shall showcase in the following. Now at this stage, a few comments
on the kinetic axion and canonical misalignment axion models
should be made. We shall mainly focus on the kinetic axion model,
but it is worth presenting both the kinetic and canonical
misalignment axion models for completeness. Both the canonical
misalignment and kinetic axion models belong to the class of
misalignment axion models \cite{Marsh:2015xka,Co:2019jts}, in
which case the pre-inflationary Peccei-Quinn $U(1)$ symmetry is
basically broken during the inflationary era and the axion field
has an initial misalignment from the minimum of its potential. In
this misalignment position, its vacuum expectation value is quite
large $\phi_i\sim \theta f_a$, with $f_a$ being the axion decay
constant, with $f_a>10^{9}\,$GeV, and $\theta$ is the misalignment
angle, taking values in the range $0<\theta <1$. The general axion
potential, after the breaking of the primordial Peccei-Quinn
$U(1)$ symmetry has the form,
\begin{equation}\label{axionpotentnialfull}
V_a(\phi )=m_a^2f_a^2\left(1-\cos \left(\frac{\phi}{f_a}\right)
\right)\, .
\end{equation}
When $\phi/f_a<1$, we can approximate the misalignment axion
potential in the following way,
\begin{equation}\label{axionpotential}
V_a(\phi )\simeq \frac{1}{2}m_a^2\phi^2\, .
\end{equation}
Now regarding the initial kinetic energy of the axion, there are
two mainstream models, the canonical misalignment axion
\cite{Marsh:2015xka} and the kinetic axion model
\cite{Co:2019jts}. In the former case, the axion rolls from its
misalignment position towards the minimum of the potential with
zero kinetic energy, and in the latter scenario the axion has a
large kinetic energy. In the canonical misalignment model, when
the axion reaches the potential minimum, and exactly when the
Hubble rate satisfies $H\sim m_a$, the axion commences
oscillations and its energy density $\rho_a$ redshifts as cold
dark matter $\rho_a\sim a^{-3}$. In the kinetic axion case, the
oscillations era starts at a much more later time compared to the
canonical misalignment because the kinetic energy of the axion
allows it to surpass the potential minimum and climbs up the
potential, as it is shown in Fig.
\ref{pictorialrepresentationaxion}. In this case, the inflationary
era lasts for more $e$-foldings
\cite{Oikonomou:2022tux,Oikonomou:2022ela} and the reheating
temperature must be smaller compared to the canonical misalignment
one.
Since the constraint on the propagation velocity of gravitational
waves and the mechanism for the kinetic axion have been briefly
discussed, let us proceed with the inflationary phenomenology.
According to the gravitational action (\ref{action}) and the
constraint (\ref{difeq}), the equations of motion read,
\begin{equation}
\centering
\label{motion1}
\frac{3FH^2}{\kappa^2}=\frac{1}{2}\dot\phi^2+V+24\dot\xi H^3+\frac{FR-f}{2\kappa^2}-\frac{3H\dot F}{\kappa^2}\, ,
\end{equation}
\begin{equation}
\centering
\label{motion2}
-\frac{2F\dot H}{\kappa^2}=\dot\phi^2-16\dot\xi H\dot H+\frac{\ddot F-H\dot F}{\kappa^2}\, ,
\end{equation}
\begin{equation}
\centering
\label{motion3}
\ddot\phi+3H\dot\phi+V'+\xi'\mathcal{G}=0\, ,
\end{equation}
where for simplicity we introduced the notation,
$F=\frac{df}{dR}$. In order to simplify the analysis, we introduce
two dimensionless auxiliary variables
$x=\frac{\kappa^2\dot\phi^2}{6FH^2}$ and
$y=-\frac{4\kappa^2\dot\xi H}{F}$ which participate in the first
two equations. In order to proceed, and by following
\cite{Oikonomou:2022tux}, the aforementioned parameters have a
negligible contribution compared to the rest terms and, due to the
fact that the axion mass is quite small and the potential and
kinetic terms of the axion are inferior to the modified gravity
terms. Thus, one finds that the background field equations can be
approximated as,
\begin{equation}
\centering
\label{motion4}
3FH^2\simeq\frac{FR-f}{2}-3H\dot F\, ,
\end{equation}
\begin{equation}
\centering
\label{motion5}
-2F\dot H\simeq\ddot F-H\dot F\, ,
\end{equation}
\begin{equation}
\centering \label{motion6}
\ddot\phi+3H\dot\phi+24\xi'H^4\bigg(1+\frac{\dot
H}{H^2}\bigg)\simeq 0\, .
\end{equation}
In order to proceed, the $f(R)$ gravity shall be chosen to be a
popular model, namely the well-known $R^2$ model,
\begin{equation}
\centering \label{fR} f(R)=R+\frac{R^2}{6M^2}\, ,
\end{equation}
where $M$ is an arbitrary mass scale determined by standard
phenomenology to be $M= 1.5\times
10^{-5}\left(\frac{N}{50}\right)^{-1}M_P$ \cite{Appleby:2009uf},
therefore for $N\sim 60$, $M$ is of the order $M\simeq
3.04375\times 10^{22}$eV. The above simplifications in the
background field equations are well justified, since, $\langle
\phi \rangle =f_a\simeq \mathcal{O}(10^{9})$GeV and we shall take
approximately $m_a\simeq \mathcal{O}(10^{-10})$eV. Thus,  the
potential term is of the order $\kappa^2V(\phi_i)\sim
\mathcal{O}(8.41897\times 10^{-30})$eV$^{2}$, and the terms $R$
and $R^2$ are of the order, $R\sim 1.2\times
\mathcal{O}(10^{45})$eV$^2$ and also $R^2/M^2\sim
\mathcal{O}(1.55\times 10^{45})$eV$^2$ for a low-scale
inflationary scenario $H_I\sim \mathcal{O}(10^{13})\,$GeV. As the
Friedmann and Raychaudhuri equations suggest, the $f(R)$ gravity
term dominate and thus these determine the Hubble rate. Additional
inclusions that dominate in the limit $R\to 0$, and thus unify
both early and late time, are also plausible scenarios
\cite{Odintsov:2020nwm} but we shall not consider these here. In
consequence, the Hubble rate expansion, due to the fact that
inflation is described by a quasi-de Sitter expansion, it should
scale linearly with time and thus the Hubble rate reads,
\begin{equation}
\centering
\label{Hubble}
H(t)=H_I-\frac{M^2}{6}t\, ,
\end{equation}
where $H_I$ is the dominant part of the Hubble rate, the scale of
inflation basically, and is assumed to be of order
$H_I\sim\mathcal{O}(10^{13})\,$GeV, so we assume a low-scale
inflationary scenario. Concerning the scalar field, it becomes
clear that its evolution is affected by the Gauss-Bonnet density.
Let us see how one can study inflation. We define the slow-roll
indices \cite{Hwang:2005hb},
\begin{align}
\centering
\label{indices}
\epsilon_1&=-\frac{\dot H}{H^2}\, ,&\epsilon_2&=\frac{\ddot\phi}{H\dot\phi}\, ,&\epsilon_3&=\frac{\dot F}{2HF}\, ,&\epsilon_4&=\frac{\dot E}{2HE}\, ,&\epsilon_5&=\frac{\dot Q_t}{2HQ_t}\, ,
\end{align}
where $E=\frac{F}{\kappa^2}\bigg[1+\frac{3(\dot
F+\kappa^2Q_a)^2}{2\kappa^4Q_t\dot\phi^2}\bigg]$, $Q_a=-8\dot\xi
H$ and $Q_t=\frac{F}{\kappa^2}+\frac{Q_a}{H}$ are auxiliary
parameters. In principle, the aforementioned indices should not be
labelled as slow-roll indices given that their numerical value is
not necessarily of order $\mathcal{O}(10^{-3})$ and below. As an example,
consider the case of a linear Gauss-Bonnet scalar coupling
function \cite{Oikonomou:2020oil}. By imposing the constraint on
the propagation velocity of tensor perturbations (\ref{difeq}), it
can easily be inferred that the second index becomes identically
equal to unity. In a sense, the large value is not an issue,
provided that compatible with the observations results can indeed
be extracted. Now obviously, one can see that indices $\epsilon_1$
and $\epsilon_3$ are affected solely by the $f(R)$ part whereas
the rest carry information about string corrections. Indeed, by
following the standard approach for the $R^2$ model, one can
easily see that during the first horizon crossing in which we are
interested in,
\begin{align}
\centering
\label{slowrollindices}
\epsilon_1&=\frac{1}{2N+1}\, ,&\epsilon_3&=-\epsilon_1+3\epsilon_1^2\, ,
\end{align}
where $N$ denotes the $e$-foldings number and is considered to be
near $N\sim60$. For the second index, by combining equations
(\ref{difeq2}) and (\ref{motion6}), it turns out that,
\begin{equation}
\centering
\label{index2}
\epsilon_2=-1+2\sqrt{1+6\xi''H^2(1-\epsilon_1)}\, ,
\end{equation}
where we can see that it depends both on the $f(R)$ part, due to
the fact that the Hubble rate expansion carries information about
the mass scale $M$, and also on the choice of the Gauss-Bonnet
scalar coupling function. Furthermore, while the aforementioned
index has no issue with the choice of the linear coupling, this
does not apply to the solution of $\dot\phi$ which in this case
reads,
\begin{equation}
\centering
\label{dotphi}
\dot\phi=\frac{2H\xi'}{\xi''}\bigg[1-\sqrt{1+6\xi''H^2(1-\epsilon_1)}\bigg]\, .
\end{equation}
The linear coupling is a special case and will be studied on its
own in the following however no matter the coupling, the
gravitational wave constraint reduces the degrees of freedom and
now the continuity equation is altered to a first order
differential equation. This expression is quite interesting in the
kinetic axion model as it \emph{implies that at the end of the
inflationary era, $\dot\phi=0$ regardless of the coupling which is
used}, therefore the \emph{axion dynamics are greatly affected}.
Regarding the rest indices, one can easily show that without
performing any approximations, they can be written with respect to
the previously defined parameters
$x=\frac{\kappa^2\dot\phi^2}{6FH^2}$ and
$y=\frac{\kappa^2Q_a}{2FH}$ as well as the rest indices as,
\begin{equation}
\centering \label{index4}
\epsilon_4=\epsilon_3+\bigg[1-\frac{x(1+2y)}{x(1+2y)+(\epsilon_3+y)^2}\bigg]\bigg[\frac{-\epsilon_1\epsilon_3+y(1-2\epsilon_1)}{\epsilon_3+y}-\epsilon_2-\epsilon_5\bigg]\,
,
\end{equation}
and,
\begin{equation}
\centering
\label{index5}
\epsilon_5=\frac{\epsilon_3+y(1-\epsilon_1)}{1+2y}\, .
\end{equation}
As a check, one can show that in the limit of $y\to 0$, meaning
that when string corrections are neglected, the respective
expressions match the results of Ref \cite{Oikonomou:2022tux}. As
a final note, we mention that the observed indices that we are
interested in, namely the scalar and tensor spectral indices of
primordial curvature perturbations along with the tensor-to-scalar
ratio, they can be computed by making use of the numerical value
of the $\epsilon_i$ indices during the first horizon crossing. In
the end, we find that from their definition \cite{Hwang:2005hb},
the observed indices can be written as,
\begin{align}
\centering
\label{observables}
n_\mathcal{S}&=1-\frac{2(2\epsilon_1+\epsilon_2-\epsilon_3+\epsilon_4)}{1-\epsilon_1}\, ,&r&=16\bigg|\bigg(3\epsilon_1+2y\bigg)\frac{\epsilon_1c_\mathcal{A}^3}{1+2y}\bigg|\, ,&n_\mathcal{T}&=-\frac{2(\epsilon_1+\epsilon_5)}{1-\epsilon_1}\, ,
\end{align}
where $c_\mathcal{A}$ stands for the propagation velocity of
scalar perturbations and in this approach is equal to,
\begin{equation}
\centering
\label{cA}
c_\mathcal{A}^2=1-\frac{4y\epsilon_1(\epsilon_3+y)}{3(x(1+2y)+(\epsilon_3+y)^2)}\, ,
\end{equation}
where due to the fact that $\epsilon_1\ll 1$, $\epsilon_3\ll 1$
and $y\ll 1$, it is expected that the sound wave velocity is
approximately equal to unity. Here, due to the fact that index
$\epsilon_5$ participates in the tensor spectral index, it could
be possible to obtain a blue-tilted tensor spectral index once the
condition $\epsilon_5<-\epsilon_1$ is satisfied. In reality, this
is impossible for the model at hand due to the fact that index
$\epsilon_5$ carries information about strings through the
dimensionless parameter $y$. Such parameter was already
encountered in the background equations
(\ref{motion1})-(\ref{motion2}) where, by making the assumption
that is inferior compared to the $f(R)$ contribution, it was
neglected. In turn, this implies that the overall phenomenology is
viable if $y\ll\epsilon_3$ but if that is the case, then no matter
the choice of Gauss-Bonnet coupling, the tensor spectral index and
the tensor-to-scalar ratio should be identical to the results of
the vacuum $R^2$. Hence, the kinetic axion model in the presence
of a Gauss-Bonnet term can only affect the scalar spectral index
at best provided that the second index is of order
$\mathcal{O}(1)$. Indeed we shall showcase this explicitly in the
following models. In total, a blue tilted tensor spectral index
could be generated if the following conditions are met,
\begin{align}
\centering
\label{positivent}
y&\leq-\frac{3\epsilon_1^2}{1+\epsilon_1}\, ,& 1+2y&>0\, ,
\end{align}
but the dominance of the $f(R)$ will never allow this conditions
to be true since $F$ suppresses $y$. This is because for $N\sim
60$, $3\epsilon_1^2\sim\mathcal{O}(10^{-4})$ however if
$1>-y>\mathcal{O}(10^{-4})$, it should participate in the
simplified background equations (\ref{motion4})-(\ref{motion5})
therefore, in this context, a viable inflationary era manages to
manifest only red-tilted tensor spectral indices. If however, the
scalar field were to be more dominant than the $f(R)$ part,
something which may be feasible for quite large mass scales $M$,
then the results could differ. This case shall not be studied
here.

As a final note regarding the expressions of the spectral indices,
one may argue that having quite large values for the indices
$\epsilon_2$ and in turn $\epsilon_4$ may be an issue with these
expressions, since they were derived by making certain
assumptions. While this is indeed the case, the above expressions
can be used without any problem, because in order to derive such
formulas, the following inequalities must be respected,
\begin{align}
\centering
\label{indequalities}
-\epsilon_1-2(\epsilon_2-\epsilon_3+\epsilon_4)&\leq3\, ,&\epsilon_1-2\epsilon_3&\leq3\, ,
\end{align}
which, due to the fact that $\epsilon_2+\epsilon_4$ is quite small
and $\epsilon_1$, $\epsilon_3$ scale with $\frac{1}{N}$, the
inequalities are indeed respected. Another issue that needs to be
addressed is the condition $\dot\epsilon_i=0$ which was assumed in
order to derive the spectral indices. While the second index
satisfies this identically in the case of a linear Gauss-Bonnet
scalar coupling function, since in this case $\epsilon_2=1$, the
rest indices evolve dynamically. In Ref. \cite{Oikonomou:2020krq}
it was shown that the condition $\dot\epsilon_i=0$ is not actually
needed but it is required so that the derivative varies slowly,
something which is indeed the case since
$\frac{\dot\epsilon_1}{H\epsilon_1}=2\epsilon_1$ and the rest can
be derived based on this. This applies to the case of an arbitrary
Gauss-Bonnet scalar coupling function as well.

\section{Inflationary Phenomenology and Cosmological Viability of Specific Models}

In this section we shall briefly showcase the results that are
produced for two models of interest. The reason why these models
were selected is because in the absence of an $f(R)$ gravity, it
was explicitly shown that the constraint on the propagation
velocity of tensor perturbations and these models, are at variance
as the results produced were not in agreement with observations.
Therefore, it is reasonable to return to these models and try to
see whether these can be rectified by the inclusion of an $R^2$
term that dominates the cosmic evolution.

\subsection{The Choice of a Linear non-minimal Gauss-Bonnet Coupling
Function}

We commence by studying the inflationary dynamics of the linear
non-minimal Gauss-Bonnet coupling,
\begin{equation}
\centering
\label{xi1}
\xi(\phi)=\frac{\phi}{f}\, ,
\end{equation}
where $f$ is an arbitrary parameter with eV mass dimensions, not
to be confused with the axion decay constant. As mentioned before,
this is an interesting model, due to the fact that it predicts
that under the assumption $c_\mathcal{T}^2=1$, the constant-roll
condition $\epsilon_2=1$ emerges naturally. In Ref.
\cite{Oikonomou:2020oil}, it was shown that the constant-roll
condition is in fact so dominant that it spoils the viability of
the scalar spectral index, however in the present context, the
inclusion of the $f(R)$ gravity results in a smooth cancellation
of the dominant $\epsilon_2$ terms, thus rendering the model
viable. Now for a linear coupling, Eq. (\ref{motion6}) suggests
that,
\begin{equation}
\centering \label{dotphi} \dot\phi=-\frac{6H^3(1-\epsilon_1)}{f}\,
.
\end{equation}
This is the general solution assuming that the contribution from
the scalar potential is subleading. It is interesting to note that
while the axion has a superior kinetic term at first, when
inflation ceases and $\epsilon_1$ becomes equal to unity, it can
easily be inferred that \emph{$\dot\phi=0$}. Now this implies that
the potential dominates and therefore, due to the inclusion of the
Gauss-Bonnet density, no stiff matter is predicted. This is a
major outcome of the present work because the stiff kination era
present in the kinetic $f(R)$ gravity case, is absent once the
non-minimal string originating Gauss-Bonnet term is absent.
\begin{figure}[h!]
\centering
\includegraphics[width=15pc]{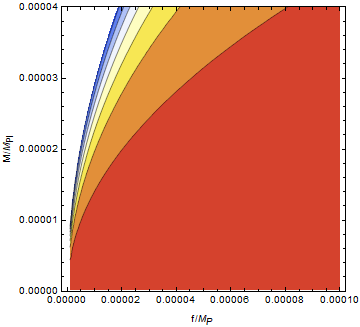}
\includegraphics[width=3.2pc]{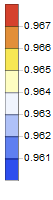}
\caption{Scalar spectral index as a function parameter $f$ and
mass scale $M$.} \label{plot1}
\end{figure}
In fact, even if one were to keep the potential, the contribution
at $\epsilon_1=1$ would be $\dot\phi=-\frac{m_\alpha^2\phi}{4H}$
and since $H\gg m_\alpha$ initially, the kinetic term becomes
inferior at that point. This is in the antipode of the result
obtained in Ref. \cite{Oikonomou:2022tux}, and this behavior
occurs only because the continuity equation can be solved
algebraically with respect to $\dot\phi$ in the case at hand, due
to the inclusion of the Gauss-Bonnet density and the constraint
(\ref{difeq}). We shall leave this issue for the time being and
focus on the inflationary era only.

\begin{figure}[h!]
\centering
\includegraphics[width=25pc]{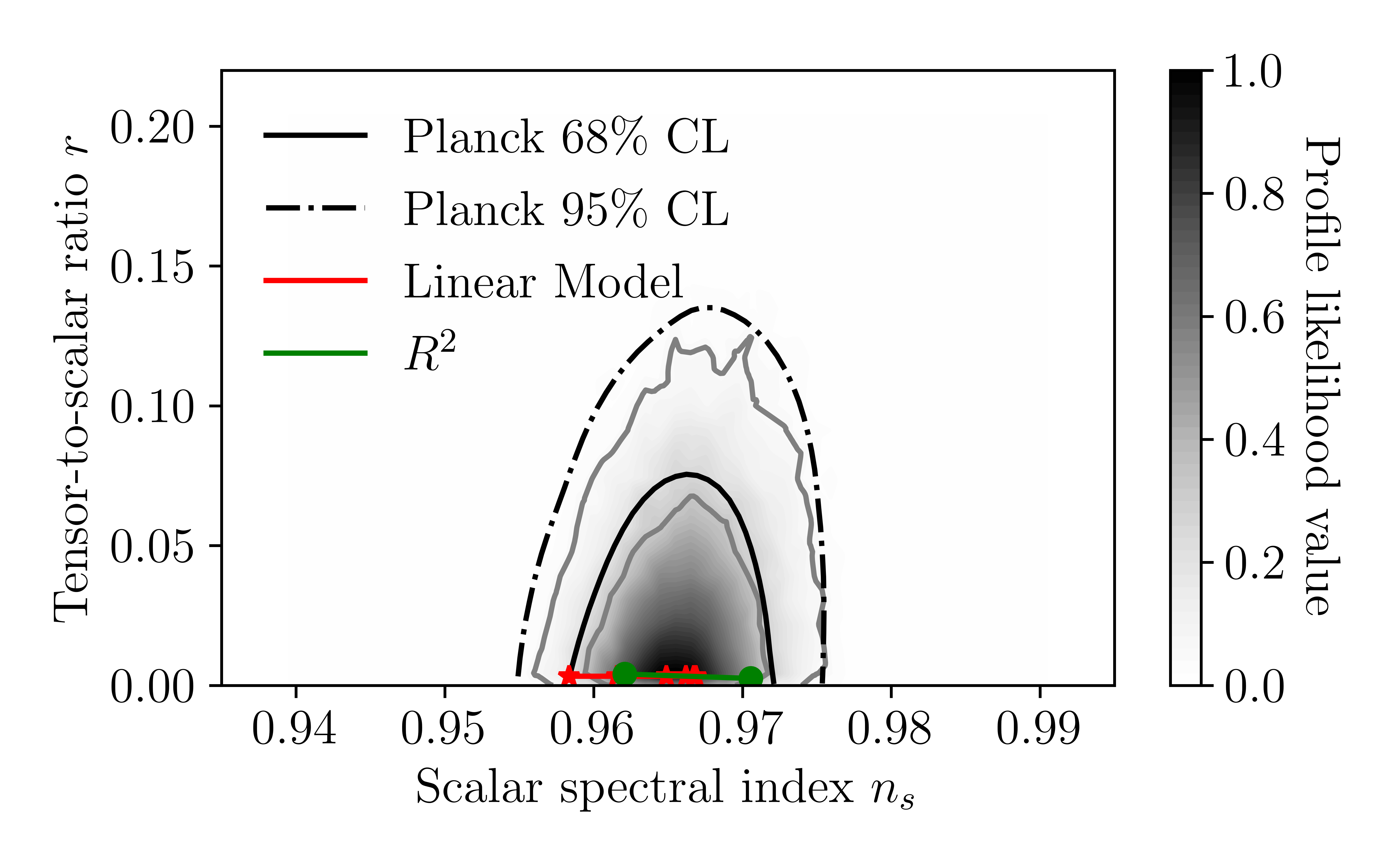}
\caption{Comparison of the linear non-minimal Gauss-Bonnet
coupling model to the $R^2$ model and the latest Planck likelihood
curves.} \label{planckcomplinear}
\end{figure}

Since $\dot\phi$ has been extracted, a simple designation of the
free parameters of the model suffices in order to derive results.
In particular, since the Hubble rate expansion in the first
horizon crossing is given by the relation,
\begin{equation}
\centering
\label{hubble}
H=\frac{M}{\sqrt{6}}\sqrt{2N+1}\, ,
\end{equation}
we find that,
\begin{equation}
\centering
\label{dotphi2}
\dot\phi=-\frac{2M^3(N+\frac{1}{2})^{\frac{3}{2}}}{f\sqrt{3}}\, ,
\end{equation}
therefore parameters $x$ $y$, $\epsilon_i$ and in consequence the
observed indices are completely specified by $N$, $f$ and $M$. From a
numerical standpoint, for $N=60$, $M=1.25\times 10^{-5}M_P$ and
$f=10^{-5}M_P$, one finds that $n_\mathcal{S}=0.96686$,
$n_\mathcal{T}=-0.000413283$ and $r=0.00327847$ which is in
agreement with the latest Planck data \cite{Planck:2018vyg}. As
expected, the results do not differ so much from the vacuum $R^2$
model exactly because string corrections and the scalar field were
assumed to be inferior to the $f(R)$ part. This can also be seen
in Fig. \ref{planckcomplinear} where we confront the linear model
at hand with the latest Planck likelihood curves and the pure
$R^2$ model results are also included.
This can be inferred from the numerical values of $x$ and $y$
which read $7\times10^{-9}$ and $3\times10^{-8}$ respectively,
therefore neglecting the kinetic term and the string corrections
in the background equations, this is justified. It is also
interesting to mention that the above results are independent of
the initial value $\phi_i$ of the scalar field, and thus the
results are valid for a plethora of values for $\phi_i$ and
$f_\alpha$. The only condition in order for the approximation in
(\ref{axionpotential}) to be valid is to demand that
$\phi_i\leq\frac{f_\alpha}{10}$, therefore if $f$ in the above
model is identified as the axion decay constant such that
$f_\alpha\sim\mathcal{O}(10^{13})$GeV then
$\phi_i\leq\mathcal{O}(10^{12})$GeV. According to Fig.
\ref{plot1}, $f$ cannot be decreased any further since the scalar
spectral index becomes incompatible with observations.

As a final note, it should be mentioned that the constant-roll
condition for the scalar field $\ddot\phi=H\dot\phi$, due to the
constraint on the propagation velocity of tensor perturbations, is
connected to the production of scalar non-Gaussianities in the
CMB. In principle, a detailed analysis should be made however, due
to the fact that the $f(R)$ part dominates, one can expect that no
matter the numerical value of index $\epsilon_2$ during the first
horizon crossing where $k=H a$, the equilateral nonlinear term
$f_{NL}^{eq}$ should be at leading order approximately equal to
\cite{Maldacena:2002vr}
\begin{equation}
\centering
\label{fNL}
f_{NL}^{eq}=\frac{5}{12}(1-n_\mathcal{S})\, ,
\end{equation}
where according to the previous results, it should be
approximately of order $\mathcal{O}(10^{-2})$. Overall, the
kinetic axion scalar field cannot enhance the vacuum $R^2$ result,
as it is subleading.

Now the novel outcome of this model is that the kinetic axion
mechanism for the string-corrected $R^2$ gravity model does not
produce a kination era after the end of inflation. As we showed,
the condition $\dot{\phi}\sim 0$ is imposed by the physics of the
problem in the case at hand, therefore at the end of inflation,
the effective equation of state of the axion is basically a nearly
de-Sitter slightly turned to quintessential one. Now the physics
in this scenario could be interesting, since the Universe may be
overwhelmed by an early dark energy era caused by the kinetic
axion which behaves as a slow-roll scalar at the end of inflation.
This behavior is entirely caused by the extra Gauss-Bonnet term.
One thing is certain for sure, the axion oscillations must be
significantly delayed in this model, since the condition
$\dot{\phi}\simeq 0$ clearly indicates that the evolution of the
axion down to the minimum of its potential is delayed, and the
axion slowly-rolls down to its minimum in a modulated way
controlled by the Gauss-Bonnet coupling. The axion in this
scenario is not driving the initial inflationary era, but it seems
that it may control the post-$f(R)$ gravity inflation cosmological
era. The vacuum fluctuations of the $R^2$ term may not be able to
initiate the reheating era. Therefore, what we may are facing in
this case is basically a second short slow-roll era controlled by
the axion prior the Hubble rate reaches the value $m_a\sim H$. It
is thus a physical situation where the $R^2$ fluctuations are
competing a slow-roll axion, but there is a caveat in this
scenario, mainly the fact that the energy density of the axion
$\rho_a\sim a^{-3(1+w)}$ for a nearly de-Sitter era is almost
constant $\rho_a\sim \mathrm{const}$, while in the kinetic axion
case $\rho_a\sim a^{-6}$ and for a purely matter dominated
post-inflationary era $\rho_a\sim a^{-3}$. Hence, it is apparent
that the contribution of the axion to the $R^2$ reheating process
is comparable to $\kappa^2 \rho_a\sim \kappa^2$ thus the standard
$R^2$ reheating occurs. In this case however, the axion
oscillations are somewhat delayed, but slightly.

\subsection{The Choice of an Exponential non-minimal Gauss-Bonnet Coupling
Function}

The second model that shall be discussed is the case in which the
Gauss-Bonnet coupling has an exponential form,
\begin{equation}
\centering \label{xi2} \xi(\phi)=\e^{-\frac{\phi}{f}}\, .
\end{equation}
This model was studied in Ref. \cite{Odintsov:2020sqy} and as it
was shown, the exponential model resulted, depending on the value
of $f$, to either eternal or no inflation at all. Therefore, it is
intriguing to study the exponential model in this case in order to
examine under which circumstances it can result to a viable
inflationary era. Now, since the Gauss-Bonnet coupling is not
linear, the following set of equations need to be used,
\begin{align}
\centering
\label{set}
\dot\phi&=\frac{2H\xi'}{\xi''}\bigg[1-\sqrt{1+6\xi''H^2(1-\epsilon_1)}\bigg]\, ,&\epsilon_2&=-1+2\sqrt{1+6\xi''H^2(1-\epsilon_1)}\, ,
\end{align}
where due to the fact that the factor $1-\epsilon_1$ appears, the
time derivative at the end of inflation is identically equal to
zero. Similarly as in the previous case, this outcome suggests
that the kinetic term is not dominant compared to the potential,
therefore the intermediate stiff matter era that emerged in Ref.
\cite{Oikonomou:2022tux} is avoided in this model too. In other
words, the inclusion of the Gauss-Bonnet density in Eq.
(\ref{action}) suggests that it cannot result in a reduction of
the tensor-to-scalar ratio due to the fact that the inflationary
era cannot be prolonged any further. Even if the potential is
present in the continuity equation, then the solution of
(\ref{motion3}) accounting for the constraint (\ref{difeq})
suggests that,
\begin{equation}
\centering
\label{puredotphi}
\dot\phi=\frac{2H\xi'}{\xi''}\bigg[1-\sqrt{1+6H^2\xi''(1-\epsilon_1)+\frac{V'\xi''}{4\xi'H^2}}\bigg]\, ,
\end{equation}
therefore for $\epsilon_1=1$, one can see that the ratio between
the kinetic term and the scalar potential of the axion is,
\begin{equation}
\centering
\label{ratio}
\frac{\dot\phi^2}{2V}=\bigg(\frac{2\xi'}{\xi''\phi}\bigg)^2\bigg(\frac{H}{m_\alpha}\bigg)^2\bigg[1-\sqrt{1+\frac{\phi\xi''}{4\xi'}\bigg(\frac{m_\alpha}{H}\bigg)^2}\bigg]^2\, .
\end{equation}
Provided that at the end of inflation, $\frac{H}{m_\alpha}\gg1$,
then by performing a Taylor expansion one finds that,
\begin{equation}
\centering
\label{ratio2}
\frac{\dot\phi^2}{2V}\simeq\bigg(\frac{m_\alpha}{4H}\bigg)^2\, ,
\end{equation}
and thus, the kinetic term is inferior therefore the model does
not predict any increase in the duration of inflation, exactly as
was the case with the linear Gauss-Bonnet coupling. Therefore,
this result is the same regardless of the coupling chosen.
\begin{figure}[h!]
\centering
\includegraphics[width=15pc]{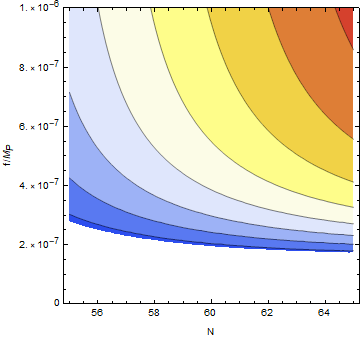}
\includegraphics[width=3.2pc]{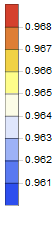}
\includegraphics[width=15pc]{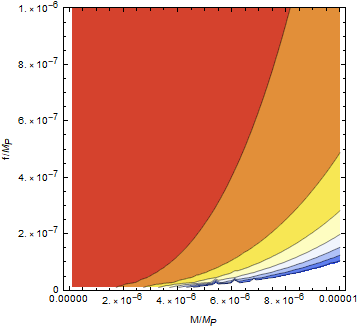}
\includegraphics[width=2.9pc]{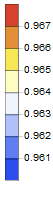}
\caption{Spectral index $n_\mathcal{S}$ depending on parameters
$f$ and $N$ on the left and $f$ and $M$ on the right.}
\label{plot2}
\end{figure}
Let us now proceed with the numerical results of the model at
hand. By using similar values as in the linear coupling such as
$N=60$ and $M=1.25\times10^{-5}M_P$, for the case of
$f=10^{-6}M_P$ and $\phi_i=10^{-10}M_P$ one finds that
$n_\mathcal{S}= 0.965791$, $n_\mathcal{T}=-0.000413309$ and
$r=0.00327848$ which are obviously in agreement with Planck data.
As expected, since $x=1.5\times 10^{-10}$ and $y=4\times10^{-8}$,
the scalar field is subleading in background equations and the
tensor spectral index and tensor-to-scalar ratio are equivalent to
their vacuum $R^2$ counterparts and only the scalar spectral index
is affected to some degree, due to the small value of $f$. It
should be stated that such value increases extremely the second
and fourth index as now $\epsilon_2=273.994$ and
$\epsilon_4=-274.001$ however in order to have correct results, we
are interested in their sum and their slow-variation as it was
mentioned previously. Obviously, larger values of $f$ decrease the
aforementioned indices but do not influence the scalar spectral
index, see Fig.\ref{plot2} for further details on the scalar
spectral index. Finally, since the initial value of the scalar
field was assumed to be approximately
$\phi_i\sim\mathcal{O}(10^8)$ GeV, the approximation on the
canonical potential (\ref{axionpotential}) applies if the axion
decay constant is $f_\alpha\geq\mathcal{O}(10^{9})$GeV.

As a general note, the inclusion of additional string terms does
not seem to influence the overall phenomenology. A
back-of-the-envelope calculation for the case of an additional
string correction,
\begin{equation}
\centering
\label{exampleaction2}
\mathcal{S}=\int d^4x\sqrt{-g}\bigg(\frac{f(R)}{2\kappa^2}-\frac{1}{2}g^{\mu\nu}\nabla_\mu\phi\nabla_\nu\phi-V(\phi)-\xi(\phi)\bigg(\mathcal{G}+cG^{\mu\nu}\nabla_\mu\phi\nabla_\nu\phi\bigg)\bigg)\, ,
\end{equation}
that also affects tensor perturbations suggests that if the
constraint $c_\mathcal{T}^2=1$ is once again imposed such that
$\ddot\xi=H\dot\xi-\frac{c\xi\dot\phi^2}{4}$, the numerical
results are not affected for similar set of values for the free
parameters and $\dot\phi^2(t_{end})\ll V_{end}$ once again.

\section{Phase Space Analysis of the Model}

In the last section of this paper we shall briefly discuss the
phase space of the proposed kinetic axion $f(R)$ gravity model. By
including perfect matter fluids, including radiation and dark
matter fluids, the gravitational action reads,
\begin{equation}
\centering
\label{action2}
\mathcal{S}=\int d^4x\sqrt{-g}\bigg(\frac{f(R)}{2\kappa^2}-\frac{1}{2}g^{\mu\nu}\nabla_\mu\phi\nabla_\nu\phi-V(\phi)-\xi(\phi)\mathcal{G}+\mathcal{L}_{matter}\bigg)\, ,
\end{equation}
and in consequence, the background equations (\ref{motion1}) and
(\ref{motion2}) are rewritten as,
\begin{equation}
\centering
\label{motion7}
\frac{3FH^2}{\kappa^2}=\rho+\frac{FR-f}{2\kappa^2}-\frac{3H\dot F}{\kappa^2}+\frac{1}{2}\dot\phi^2+V\, ,
\end{equation}
\begin{equation}
\centering \label{motion8} -\frac{2F\dot
H}{\kappa^2}=\rho+P+\frac{\ddot F-H\dot F}{\kappa^2}+\dot\phi^2\,
,
\end{equation}
where $\rho$ and $P$ are the total energy density and pressure of
the perfect matter fluids respectively. With regard to the latter,
we shall assume that the axion is the sole dark matter component
and that other extra dark matter components are not present. In
order to perform an autonomous dynamical analysis, the
Gauss-Bonnet scalar coupling function shall remain unspecified and
the rest functions of the model will follow a power-law form,
\begin{align}
\centering
\label{functions}
f(R)&=R+AR^n\, ,&V(\phi)&=V_0(\kappa\phi)^m\, ,
\end{align}
for the sake of generality however we shall limit our results in
the case of $n=2=m$ where in turn $A=\frac{1}{6M^2}$ and
$V_0=\frac{m_\alpha^2}{2\kappa^2}$. In addition, let us define the
following dynamical variables,
\begin{align}
\centering
\label{variables}
x&=\frac{\kappa^2\dot\phi^2}{6FH^2}\, ,&y&=\frac{\kappa^2V}{3FH^2}\, ,&z&=\frac{8\kappa^2\dot\xi H}{F}\, ,&u&=-\frac{\dot F}{HF}\, ,&v&=-\frac{f}{6FH^2}\, ,&w&=\frac{R}{6H^2}\, ,&s&=\frac{\dot V}{HV}\, ,&p&=\frac{\kappa^2\rho_r}{3FH^2}\, ,&q&=\frac{\kappa^2\rho_m}{3FH^2}\, ,
\end{align}
where $p$ and $q$ should not be mistaken with $\Omega_r$ and
$\Omega_m$ due to the $F$ part in the denominator. In consequence,
equations (\ref{motion7}) and (\ref{motion3}) are rewritten as,
\begin{equation}
\centering
\label{friedmannconstraint}
x+y+z+u+v+w+p+q=1\, ,
\end{equation}
\begin{equation}
\centering
\label{ddotphi}
\frac{\ddot\phi}{H\dot\phi}=-3-\frac{s}{2}\frac{y}{x}-\frac{z}{2x}(w-1)\, .
\end{equation}
Now by making use of the $e$-foldings number through the
differential equation $\frac{d}{dN}=\frac{1}{H}\frac{d}{dt}$, the
following set of differential equations is produced,
\begin{equation}
\centering
\label{dx}
\frac{dx}{dN}=x\bigg[-6-\frac{y}{x}-\frac{z}{x}(w-1)+u-2(w-2)\bigg]\, ,
\end{equation}
\begin{equation}
\centering
\label{dy}
\frac{dy}{dN}=y\bigg[s+u-2(w-2)\bigg]\, ,
\end{equation}
\begin{equation}
\centering
\label{dz}
\frac{dz}{dN}=z\bigg[w-1+u\bigg]\, ,
\end{equation}
\begin{equation}
\centering
\label{du}
\frac{du}{dN}=u(u+1)+6x+4p+3q+2(w-2)(1-z-\frac{u}{2})\, ,
\end{equation}
\begin{equation}
\centering
\label{dv}
\frac{dv}{dN}=v(u-2(w-2))+\frac{uw^2}{n(v+w)}\, ,
\end{equation}
\begin{equation}
\centering
\label{dw}
\frac{dw}{dN}=-w\bigg[\frac{uw}{n(v+w)}+2(w-2)\bigg]\, ,
\end{equation}
\begin{equation}
\centering
\label{ds}
\frac{ds}{dN}=s\bigg[-3-\frac{s}{2}\frac{y}{x}-\frac{z}{2x}(w-1)-(u-2)-\frac{s}{m}\bigg]\, ,
\end{equation}
\begin{equation}
\centering
\label{dp}
\frac{dp}{dN}=p(u-2w)\, ,
\end{equation}
\begin{equation}
\centering
\label{dq}
\frac{dq}{dN}=q(u-2w+1)\, .
\end{equation}
Let us study two separate cases. Firstly, we focus on the vacuum
$R^2$ kinetic axion model, therefore the parameters $y$ and $s$
are neglected along with $p$ and $q$. Therefore, a subsystem
comprised of variables $x$ and $z$, that are connected to the
axion dynamics and the Gauss-Bonnet term, along with $u$, $v$ and
$w$, connected to the $f(R)$ part, is studied. This subsystem can
be solved relatively straightforward. So as expected, the
inclusion of the Gauss-Bonnet produces a richer phase space. As
presented in Table \ref{table1}, there exist four fixed points in
total, one of them being a de Sitter fixed point, another is
connected to matter while additionally a third to radiation
domination eras and finally, the new stable fixed point from the
Gauss-Bonnet contribution describes acceleration and since
$\omega_{eff}=-1.6667$, it predicts a phantom evolution. This is
the case of phantom dark energy that has been covered in Ref.
\cite{Caldwell:2003vq}.
\begin{table}[h!]
\centering
\begin{tabular}{|c|c|c|c|c|c|}
\hline
Fixed Point&$(x,z,u,v,w)$&Eigenvalues&Stability&q&$\omega_{eff}$\\ \hline
$P_1$&(0,0,-4,5,0)&(-6,-5,-5,4,0) &Non Hyperbolic&1&$\frac{1}{3}$\\ \hline
$P_2$&(0,0,0,-1,0)&(0,-6,1,-3,0) &Non Hyperbolic&-1&-1\\ \hline
$P_3$&($-\frac{9}{4}$,0,3,-$\frac{1}{4}$,$\frac{1}{2}$)&(6,3,3,$\frac{5}{2}$,-$\frac{3}{2}$) &Saddle&$\frac{1}{2}$&0\\ \hline
$P_4$&($-\frac{3}{8}$,-$\frac{15}{16}$,-2,-$\frac{3}{2}$,3)&(-9.42318,-5.74568,-4,-2,-0.831139) &Stable Node&-2&-$\frac{5}{3}$\\ \hline
\end{tabular}
\caption{Fixed points for the vacuum model with subleading potential.}
\label{table1}
\end{table}
Now if one includes the potential, along with variable $s$, it
turns out that no additional fixed point emerges since $y\to 0$
and only $s$ can obtain a non-zero but finite value. This however
does not affect the Friedmann constraint neither the stability of
the fixed points, it only suggests how $y$ or in other words
$\phi$ evolves. Finally, the inclusion of perfect matter fluids
increases the number of fixed points by two, as shown in Table
\ref{table2} and surprisingly, these fixed points have nothing to
do with matter or radiation. In fact, the first fixed point is
connected to a static universe as $\omega_{eff}=-\frac{1}{3}$
whereas the second describes quintessential acceleration with
$\omega_{eff}=-\frac{1}{2}$. In a sense, the $f(R)$ gravity
describes both de-Sitter and radiation domination eras, while the
kinetic term of the axion describes matter domination. This is
somewhat expected since the axion is a dark matter component.
\begin{table}
\centering
\begin{tabular}{|c|c|c|c|c|c|}
\hline
Fixed Point&$(x,y,z,u,v,w,p,q)$&Eigenvalues&Stability&q&$\omega_{eff}$\\ \hline
$P_1$&(0,0,0,-4,5,0,0,0)&(-6,-5,-5,4,4,-3,0,0) &Non Hyperbolic&1&$\frac{1}{3}$\\ \hline
$P_2$&(0,0,0,0,-1,0,0,0)&(0,0,-6,-4,-3,1,-3,0) &Non Hyperbolic&-1&-1\\ \hline
$P_3$&($-\frac{9}{4}$,0,0,3,-$\frac{1}{4}$,$\frac{1}{2},0,0$)&(6,6,3,3,3,$\frac{5}{2}$,2,-$\frac{3}{2}$) &Saddle&$\frac{1}{2}$&0\\ \hline
$P_4$&($-\frac{3}{8}$,0,-$\frac{15}{16}$,-2,-$\frac{3}{2}$,3,0,0)&(-9.42318,-8,-7,-5.74568,-4,-4,-2,-0.831139) &Stable Node&-2&-$\frac{5}{3}$\\ \hline
$P_5$&(0,0,2,-$\frac{1}{2}$,1,0,-$\frac{3}{2}$,0 )& (4,4,-2.44949,2.44949,-2,2,2,1 )&Saddle&0 &$-\frac{1}{3}$\\ \hline
$P_6$&(0,0,$\frac{3}{2}$,-$\frac{5}{8}$,$\frac{5}{4}$0,0,-$\frac{9}{8}$ )& ( -3,3,3,-2.77617,2.02617,1.75,1.5,-1)&Saddle&$-\frac{1}{4}$ &$-\frac{1}{2}$\\ \hline
\end{tabular}
\caption{Fixed point of the complete model.}
\label{table2}
\end{table}

As a final note, it should be stated that under the assumption that
$c_{\mathcal{T}}^2=1$, the constraint in Eq. (\ref{difeq2}) should
be implemented in the continuity equation of the scalar field as
well. In consequence, if it is combined with Eq. (\ref{ddotphi}),
an additional equation emerges that needs to be satisfied at all
times and reads,

\begin{equation}
\centering
\label{conteqconstraint}
2(4-\lambda)x+sy+z(w-1)=0\, ,
\end{equation}
where $\lambda=\frac{\dot\xi'}{H\xi'}$ is an additional auxiliary
parameter. This condition is respected by the fixed points
which were previously derived and for a nonzero but finite
value of $x$, which was shown to manifest previously,
parameter $\lambda$ simply obtains a specific value. The
linear Gauss-Bonnet coupling is a special case since it
suggests that $\lambda=0$ identically therefore the fixed points
$P_3$ through $P_6$ cannot satisfy the aforementioned constraint.

We need to note that in principle, in the context of our model, it
is possible to satisfy in a concrete way all the Swampland
criteria, due to the freedom offered by the $f(R)$ sector and the
Gauss-Bonnet coupling, something which is not possible in ordinary
scalar field theories
\cite{Kinney:2018nny,Kehagias:2018uem,Achucarro:2018vey}.

\section{Conclusions}

In this work we studied the effects of a non-minimal coupling of
the kinetic misalignment axion field on the inflationary era
generated by an $R^2$ gravity. In the case of kinetic axion $R^2$
gravity, the inflationary controlled by the $R^2$ gravity is
somewhat prolonged by the kinetic axion, since at the end of
inflation, the stiff axion evolution prevails the evolution, thus
the reheating era is delayed. This is due to the fact that the
kinetic axion at the end of the inflationary era dominates over
the $R^2$ fluctuations which would start the reheating process,
and in effect, the background equation of state parameter would
have value $w=1$. In turn, this affects the total number of the
$e$-foldings and hence the inflationary era is prolonged. Also in
the same context, the reheating era is somewhat shortened, thus
the Universe in this scenario would have a lower reheating
temperature compared to the canonical misalignment axion $R^2$
model. However, as we showed in this article, the non-trivial
Gauss-Bonnet coupling can have dramatic effects on the axion
itself, imposing the condition $\dot{\phi}\simeq 0$ at the end of
the inflationary era. In this case it is obvious that the kinetic
evolution of the axion is stopped at the end of inflation, and
thus the axion reaches the minimum of its potential faster.
Therefore, it starts its oscillations around its minimum in a
standard way when its mass is of the same order as the Hubble
rate, and therefore the $R^2$ fluctuations control the reheating
era. Thus in some way, the non-trivial Gauss-Bonnet coupling of
the axion, counteracts on the kination axion mechanism,
eliminating the stiff era axion evolution at the end of the
inflationary era. At a phenomenological level, the $R^2$-corrected
canonical misalignment axion model and the kinetic axion $R^2$
inflation model with Gauss-Bonnet corrections are almost
indistinguishable.

\section*{Acknowledgments}

This research has been is funded by the Committee of Science of
the Ministry of Education and Science of the Republic of
Kazakhstan (Grant No. AP14869238)


\begin{thebibliography}{99}





\bibitem{Bertone:2004pz}
  G.~Bertone, D.~Hooper and J.~Silk,
  Phys.\ Rept.\  {\bf 405} (2005) 279
  doi:10.1016/j.physrep.2004.08.031
  [hep-ph/0404175].


\bibitem{Bergstrom:2000pn}
  L.~Bergstrom,
  Rept.\ Prog.\ Phys.\  {\bf 63} (2000) 793
  doi:10.1088/0034-4885/63/5/2r3
  [hep-ph/0002126].




\bibitem{Mambrini:2015sia}
  Y.~Mambrini, S.~Profumo and F.~S.~Queiroz,
  Phys.\ Lett.\ B {\bf 760} (2016) 807
  [arXiv:1508.06635 [hep-ph]].

\bibitem{Profumo:2013yn}
  S.~Profumo,
  arXiv:1301.0952 [hep-ph].




\bibitem{Hooper:2007qk}
  D.~Hooper and S.~Profumo,
  Phys.\ Rept.\  {\bf 453} (2007) 29
  [hep-ph/0701197].



\bibitem{Oikonomou:2006mh}
V.~K.~Oikonomou, J.~D.~Vergados and C.~C.~Moustakidis,
Nucl.\ Phys.\ B {\bf 773} (2007) 19
[hep-ph/0612293].



\bibitem{Preskill:1982cy}
  J.~Preskill, M.~B.~Wise and F.~Wilczek,
  Phys.\ Lett.\  {\bf 120B} (1983) 127.
  doi:10.1016/0370-2693(83)90637-8


\bibitem{Abbott:1982af}
  L.~F.~Abbott and P.~Sikivie,
  Phys.\ Lett.\  {\bf 120B} (1983) 133.
  doi:10.1016/0370-2693(83)90638-X


\bibitem{Dine:1982ah}
  M.~Dine and W.~Fischler,
  Phys.\ Lett.\  {\bf 120B} (1983) 137.
  doi:10.1016/0370-2693(83)90639-1







\bibitem{Marsh:2015xka}
  D.~J.~E.~Marsh,
  Phys.\ Rept.\  {\bf 643} (2016) 1
  [arXiv:1510.07633 [astro-ph.CO]].


\bibitem{Chadha-Day:2021szb}
F.~Chadha-Day, J.~Ellis and D.~J.~E.~Marsh,
Sci. Adv. \textbf{8} (2022) no.8, abj3618
doi:10.1126/sciadv.abj3618 [arXiv:2105.01406 [hep-ph]].


\bibitem{Choi:2020rgn}
K.~Choi, S.~H.~Im and C.~Sub Shin,
Ann. Rev. Nucl. Part. Sci. \textbf{71} (2021), 225-252
doi:10.1146/annurev-nucl-120720-031147 [arXiv:2012.05029
[hep-ph]].


\bibitem{DiLuzio:2020wdo}
L.~Di Luzio, M.~Giannotti, E.~Nardi and L.~Visinelli,
Phys. Rept. \textbf{870} (2020), 1-117
doi:10.1016/j.physrep.2020.06.002 [arXiv:2003.01100 [hep-ph]].




\bibitem{Sikivie:2006ni}
  P.~Sikivie,
  Lect.\ Notes Phys.\  {\bf 741} (2008) 19
  [astro-ph/0610440].



\bibitem{Raffelt:2006cw}
  G.~G.~Raffelt,
  Lect.\ Notes Phys.\  {\bf 741} (2008) 51
  [hep-ph/0611350].


\bibitem{Linde:1991km}
  A.~D.~Linde,
  Phys.\ Lett.\ B {\bf 259} (1991) 38.


\bibitem{Co:2019jts}
R.~T.~Co, L.~J.~Hall and K.~Harigaya,
Phys. Rev. Lett. \textbf{124} (2020) no.25, 251802
doi:10.1103/PhysRevLett.124.251802 [arXiv:1910.14152 [hep-ph]].



\bibitem{Co:2020dya}
R.~T.~Co, L.~J.~Hall, K.~Harigaya, K.~A.~Olive and S.~Verner,
JCAP \textbf{08} (2020), 036 doi:10.1088/1475-7516/2020/08/036
[arXiv:2004.00629 [hep-ph]].



\bibitem{Barman:2021rdr}
B.~Barman, N.~Bernal, N.~Ramberg and L.~Visinelli,
[arXiv:2111.03677 [hep-ph]].




\bibitem{Marsh:2017yvc}
  M.~C.~D.~Marsh, H.~R.~Russell, A.~C.~Fabian, B.~P.~McNamara, P.~Nulsen and C.~S.~Reynolds,
  JCAP {\bf 1712} (2017) no.12,  036
  [arXiv:1703.07354 [hep-ph]].







\bibitem{Odintsov:2019mlf}
  S.~D.~Odintsov and V.~K.~Oikonomou,
  Phys.\ Rev.\ D {\bf 99} (2019) no.6,  064049
  [arXiv:1901.05363 [gr-qc]].




\bibitem{Nojiri:2019nar}
  S.~Nojiri, S.~D.~Odintsov, V.~K.~Oikonomou and A.~A.~Popov,
  Phys.\ Rev.\ D {\bf 100} (2019) no.8,  084009
  [arXiv:1909.01324 [gr-qc]].


\bibitem{Nojiri:2019riz}
S.~Nojiri, S.~D.~Odintsov and V.~K.~Oikonomou,
Annals Phys. \textbf{418} (2020), 168186
doi:10.1016/j.aop.2020.168186 [arXiv:1907.01625 [gr-qc]].

\bibitem{Odintsov:2019evb}
  S.~D.~Odintsov and V.~K.~Oikonomou,
  Phys.\ Rev.\ D {\bf 99} (2019) no.10,  104070
  [arXiv:1905.03496 [gr-qc]].






\bibitem{Cicoli:2019ulk}
  M.~Cicoli, V.~Guidetti and F.~G.~Pedro,
  arXiv:1903.01497 [hep-th].

\bibitem{Fukunaga:2019unq}
  H.~Fukunaga, N.~Kitajima and Y.~Urakawa,
  arXiv:1903.02119 [astro-ph.CO].


\bibitem{Caputo:2019joi}
  A.~Caputo,
  arXiv:1902.02666 [hep-ph].





\bibitem{maxim}

A.S.Sakharov and M.Yu.Khlopov, 
 Yadernaya Fizika (1994) V. 57, PP. 514-
516. ( Phys.Atom.Nucl. (1994) V. 57, PP. 485-487); A.S.Sakharov,
D.D.Sokoloff and M.Yu.Khlopov, 
 Yadernaya Fizika (1996) V. 59, PP. 1050-1055.
(Phys.Atom.Nucl. (1996) V. 59, PP. 1005-1010); M .Yu.Khlopov,
A.S.Sakharov and D.D.Sokoloff,
 Nucl.Phys. B (Proc. Suppl.) (1999) V. 72, 105-109.







\bibitem{Chang:2018rso}
  J.~H.~Chang, R.~Essig and S.~D.~McDermott,
  JHEP {\bf 1809} (2018) 051
  [arXiv:1803.00993 [hep-ph]].

Chang:2018rso,Irastorza:2018dyq,

\bibitem{Irastorza:2018dyq}
  I.~G.~Irastorza and J.~Redondo,
  Prog.\ Part.\ Nucl.\ Phys.\  {\bf 102} (2018) 89
  [arXiv:1801.08127 [hep-ph]].


\bibitem{Anastassopoulos:2017ftl}
  V.~Anastassopoulos {\it et al.} [CAST Collaboration],
  Nature Phys.\  {\bf 13} (2017) 584
  [arXiv:1705.02290 [hep-ex]].








\bibitem{Sikivie:2014lha}
  P.~Sikivie,
  Phys.\ Rev.\ Lett.\  {\bf 113} (2014) no.20,  201301
  [arXiv:1409.2806 [hep-ph]].






\bibitem{Sikivie:2010bq}
  P.~Sikivie,
  Phys.\ Lett.\ B {\bf 695} (2011) 22
  [arXiv:1003.2426 [astro-ph.GA]].




\bibitem{Sikivie:2009qn}
  P.~Sikivie and Q.~Yang,
  Phys.\ Rev.\ Lett.\  {\bf 103} (2009) 111301
  [arXiv:0901.1106 [hep-ph]].



\bibitem{Caputo:2019tms}
  A.~Caputo, L.~Sberna, M.~Frias, D.~Blas, P.~Pani, L.~Shao and W.~Yan,
  Phys.\ Rev.\ D {\bf 100} (2019) no.6,  063515
  [arXiv:1902.02695 [astro-ph.CO]].




\bibitem{Masaki:2019ggg}
  E.~Masaki, A.~Aoki and J.~Soda,
  arXiv:1909.11470 [hep-ph].


\bibitem{Soda:2017sce}
  J.~Soda and D.~Yoshida,
  Galaxies {\bf 5} (2017) no.4,  96.


\bibitem{Soda:2017dsu}
  J.~Soda and Y.~Urakawa,
  Eur.\ Phys.\ J.\ C {\bf 78} (2018) no.9,  779
  [arXiv:1710.00305 [astro-ph.CO]].




\bibitem{Aoki:2017ehb}
  A.~Aoki and J.~Soda,
  Phys.\ Rev.\ D {\bf 96} (2017) no.2,  023534
  [arXiv:1703.03589 [astro-ph.CO]].










\bibitem{Masaki:2017aea}
  E.~Masaki, A.~Aoki and J.~Soda,
  Phys.\ Rev.\ D {\bf 96} (2017) no.4,  043519
  [arXiv:1702.08843 [astro-ph.CO]].


\bibitem{Aoki:2016kwl}
  A.~Aoki and J.~Soda,
  Int.\ J.\ Mod.\ Phys.\ D {\bf 26} (2016) no.07,  1750063
  [arXiv:1608.05933 [astro-ph.CO]].


\bibitem{Obata:2016xcr}
  I.~Obata and J.~Soda,
  Phys.\ Rev.\ D {\bf 94} (2016) no.4,  044062
  [arXiv:1607.01847 [astro-ph.CO]].


\bibitem{Aoki:2016mtn}
  A.~Aoki and J.~Soda,
  Phys.\ Rev.\ D {\bf 93} (2016) no.8,  083503
  [arXiv:1601.03904 [hep-ph]].



\bibitem{Ikeda:2019fvj}
  T.~Ikeda, R.~Brito and V.~Cardoso,
  Phys.\ Rev.\ Lett.\  {\bf 122} (2019) no.8,  081101
  [arXiv:1811.04950 [gr-qc]].



\bibitem{Arvanitaki:2019rax}
  A.~Arvanitaki, S.~Dimopoulos, M.~Galanis, L.~Lehner, J.~O.~Thompson and K.~Van Tilburg,
  arXiv:1909.11665 [astro-ph.CO].



\bibitem{Arvanitaki:2016qwi}
  A.~Arvanitaki, M.~Baryakhtar, S.~Dimopoulos, S.~Dubovsky and R.~Lasenby,
  Phys.\ Rev.\ D {\bf 95} (2017) no.4,  043001
  [arXiv:1604.03958 [hep-ph]].

\bibitem{Arvanitaki:2014wva}
  A.~Arvanitaki, M.~Baryakhtar and X.~Huang,
  Phys.\ Rev.\ D {\bf 91} (2015) no.8,  084011
  [arXiv:1411.2263 [hep-ph]].


\bibitem{Arvanitaki:2014dfa}
  A.~Arvanitaki and A.~A.~Geraci,
  Phys.\ Rev.\ Lett.\  {\bf 113} (2014) no.16,  161801
  [arXiv:1403.1290 [hep-ph]].




\bibitem{Sen:2018cjt}
  S.~Sen,
  Phys.\ Rev.\ D {\bf 98} (2018) no.10,  103012
  [arXiv:1805.06471 [hep-ph]].









\bibitem{Cardoso:2018tly}
  V.~Cardoso, S.~J.~C.~Dias, G.~S.~Hartnett, M.~Middleton, P.~Pani and J.~E.~Santos,
  JCAP {\bf 1803} (2018) 043
  [arXiv:1801.01420 [gr-qc]].


\bibitem{Rosa:2017ury}
  J.~G.~Rosa and T.~W.~Kephart,
  Phys.\ Rev.\ Lett.\  {\bf 120} (2018) no.23,  231102
  [arXiv:1709.06581 [gr-qc]].


\bibitem{Yoshino:2013ofa}
  H.~Yoshino and H.~Kodama,
  PTEP {\bf 2014} (2014) 043E02
  [arXiv:1312.2326 [gr-qc]].



\bibitem{Machado:2019xuc}
  C.~S.~Machado, W.~Ratzinger, P.~Schwaller and B.~A.~Stefanek,
  arXiv:1912.01007 [hep-ph].




\bibitem{Korochkin:2019qpe}
  A.~Korochkin, A.~Neronov and D.~Semikoz,
  arXiv:1911.13291 [hep-ph].


\bibitem{Chou:2019enw}
  A.~S.~Chou,
  Astrophys.\ Space Sci.\ Proc.\  {\bf 56} (2019) 41.


\bibitem{Chang:2019tvx}
  C.~F.~Chang and Y.~Cui,
  arXiv:1911.11885 [hep-ph].








\bibitem{Crisosto:2019fcj}
  N.~Crisosto, G.~Rybka, P.~Sikivie, N.~S.~Sullivan, D.~B.~Tanner and J.~Yang,
  arXiv:1911.05772 [astro-ph.CO].


\bibitem{Choi:2019jwx}
  K.~Choi, H.~Seong and S.~Yun,
  arXiv:1911.00532 [hep-ph].








\bibitem{Kavic:2019cgk}
  M.~Kavic, S.~L.~Liebling, M.~Lippert and J.~H.~Simonetti,
  arXiv:1910.06977 [astro-ph.HE].


\bibitem{Blas:2019qqp}
  D.~Blas, A.~Caputo, M.~M.~Ivanov and L.~Sberna,
  arXiv:1910.06128 [hep-ph].




\bibitem{Guerra:2019srj}
  D.~Guerra, C.~F.~B.~Macedo and P.~Pani,
  JCAP {\bf 1909} (2019) no.09,  061
  [arXiv:1909.05515 [gr-qc]].




\bibitem{Tenkanen:2019xzn}
  T.~Tenkanen and L.~Visinelli,
  JCAP {\bf 1908} (2019) 033
  [arXiv:1906.11837 [astro-ph.CO]].




\bibitem{Huang:2019rmc}
  G.~Y.~Huang and S.~Zhou,
  Phys.\ Rev.\ D {\bf 100} (2019) no.3,  035010
  [arXiv:1905.00367 [hep-ph]].




\bibitem{Croon:2019iuh}
  D.~Croon, R.~Houtz and V.~Sanz,
  JHEP {\bf 1907} (2019) 146
  [arXiv:1904.10967 [hep-ph]].


\bibitem{Day:2019bbh}
  F.~V.~Day and J.~I.~McDonald,
  JCAP {\bf 1910} (2019) no.10,  051
  [arXiv:1904.08341 [hep-ph]].


\bibitem{Odintsov:2020iui}
S.~D.~Odintsov and V.~K.~Oikonomou,
EPL \textbf{129} (2020) no.4, 40001
doi:10.1209/0295-5075/129/40001 [arXiv:2003.06671 [gr-qc]].



\bibitem{Nojiri:2020pqr}
S.~Nojiri, S.~D.~Odintsov, V.~K.~Oikonomou and A.~A.~Popov,
Phys. Dark Univ. \textbf{28} (2020), 100514
doi:10.1016/j.dark.2020.100514 [arXiv:2002.10402 [gr-qc]].



\bibitem{Odintsov:2020nwm}
S.~D.~Odintsov and V.~K.~Oikonomou,
Phys. Rev. D \textbf{101} (2020) no.4, 044009
doi:10.1103/PhysRevD.101.044009 [arXiv:2001.06830 [gr-qc]].







\bibitem{Oikonomou:2020qah}
V.~K.~Oikonomou,
Phys. Rev. D \textbf{103} (2021) no.4, 044036
doi:10.1103/PhysRevD.103.044036 [arXiv:2012.00586 [astro-ph.CO]].



\bibitem{Oikonomou:2022ela}
V.~K.~Oikonomou,
EPL \textbf{139} (2022) no.6, 69004 doi:10.1209/0295-5075/ac8fb2
[arXiv:2209.08339 [hep-ph]].


\bibitem{Oikonomou:2022tux}
V.~K.~Oikonomou,
Phys. Rev. D \textbf{106} (2022) no.4, 044041
doi:10.1103/PhysRevD.106.044041 [arXiv:2208.05544 [gr-qc]].


\bibitem{Mazde:2022sdx}
K.~Mazde and L.~Visinelli,
JCAP \textbf{01} (2023), 021 doi:10.1088/1475-7516/2023/01/021
[arXiv:2209.14307 [astro-ph.CO]].




\bibitem{Chen:2022nbb}
Y.~Chen, R.~Roy, S.~Vagnozzi and L.~Visinelli,
Phys. Rev. D \textbf{106} (2022) no.4, 043021
doi:10.1103/PhysRevD.106.043021 [arXiv:2205.06238 [astro-ph.HE]].


\bibitem{Caloni:2022uya}
L.~Caloni, M.~Gerbino, M.~Lattanzi and L.~Visinelli,
JCAP \textbf{09} (2022), 021 doi:10.1088/1475-7516/2022/09/021
[arXiv:2205.01637 [astro-ph.CO]].



\bibitem{Roy:2021uye}
R.~Roy, S.~Vagnozzi and L.~Visinelli,
Phys. Rev. D \textbf{105} (2022) no.8, 083002
doi:10.1103/PhysRevD.105.083002 [arXiv:2112.06932 [astro-ph.HE]].



\bibitem{DiLuzio:2021qct}
L.~Di Luzio, J.~Galan, M.~Giannotti, I.~G.~Irastorza, J.~Jaeckel,
A.~Lindner, J.~Ruz, U.~Schneekloth, L.~Sohl and L.~J.~Thormaehlen,
\textit{et al.}
Eur. Phys. J. C \textbf{82} (2022) no.2, 120
doi:10.1140/epjc/s10052-022-10061-1 [arXiv:2111.06407 [hep-ph]].



\bibitem{Choi:2021aze}
G.~Choi, W.~Lin, L.~Visinelli and T.~T.~Yanagida,
Phys. Rev. D \textbf{104} (2021) no.10, 10
doi:10.1103/PhysRevD.104.L101302 [arXiv:2106.12602 [hep-ph]].



\bibitem{DiLuzio:2021pxd}
L.~Di Luzio, B.~Gavela, P.~Quilez and A.~Ringwald,
JHEP \textbf{05} (2021), 184 doi:10.1007/JHEP05(2021)184
[arXiv:2102.00012 [hep-ph]].



\bibitem{Bauer:2020jbp}
M.~Bauer, M.~Neubert, S.~Renner, M.~Schnubel and A.~Thamm,
JHEP \textbf{04} (2021), 063 doi:10.1007/JHEP04(2021)063
[arXiv:2012.12272 [hep-ph]].


\bibitem{Ramberg:2020oct}
N.~Ramberg and L.~Visinelli,
Phys. Rev. D \textbf{103} (2021) no.6, 063031
doi:10.1103/PhysRevD.103.063031 [arXiv:2012.06882 [astro-ph.CO]].


\bibitem{DiLuzio:2020jjp}
L.~Di Luzio, M.~Fedele, M.~Giannotti, F.~Mescia and E.~Nardi,
Phys. Rev. Lett. \textbf{125} (2020) no.13, 131804
doi:10.1103/PhysRevLett.125.131804 [arXiv:2006.12487 [hep-ph]].



\bibitem{Visinelli:2018utg}
L.~Visinelli and S.~Vagnozzi,
Phys. Rev. D \textbf{99} (2019) no.6, 063517
doi:10.1103/PhysRevD.99.063517 [arXiv:1809.06382 [hep-ph]].



\bibitem{Visinelli:2018wza}
L.~Visinelli and J.~Redondo,
Phys. Rev. D \textbf{101} (2020) no.2, 023008
doi:10.1103/PhysRevD.101.023008 [arXiv:1808.01879 [astro-ph.CO]].


\bibitem{DiLuzio:2016sbl}
L.~Di Luzio, F.~Mescia and E.~Nardi,
Phys. Rev. Lett. \textbf{118} (2017) no.3, 031801
doi:10.1103/PhysRevLett.118.031801 [arXiv:1610.07593 [hep-ph]].





\bibitem{Semertzidis:2021rxs}
Y.~K.~Semertzidis and S.~Youn,
Sci. Adv. \textbf{8} (2022) no.8, abm9928
doi:10.1126/sciadv.abm9928 [arXiv:2104.14831 [hep-ph]].




\bibitem{Buschmann:2021sdq}
M.~Buschmann, J.~W.~Foster, A.~Hook, A.~Peterson, D.~E.~Willcox,
W.~Zhang and B.~R.~Safdi,
Nature Commun. \textbf{13} (2022) no.1, 1049
doi:10.1038/s41467-022-28669-y [arXiv:2108.05368 [hep-ph]].



\bibitem{BREAD:2021tpx}
J.~Liu \textit{et al.} [BREAD],
Phys. Rev. Lett. \textbf{128} (2022) no.13, 131801
doi:10.1103/PhysRevLett.128.131801 [arXiv:2111.12103
[physics.ins-det]].



\bibitem{Hoof:2022xbe}
S.~Hoof and L.~Schulz,
[arXiv:2212.09764 [hep-ph]].


\bibitem{Li:2022pqa}
H.~J.~Li and W.~Chao,
[arXiv:2211.00524 [hep-ph]].



\bibitem{Codello:2015mba}
A.~Codello and R.~K.~Jain,
Class. Quant. Grav. \textbf{33} (2016) no.22, 225006
doi:10.1088/0264-9381/33/22/225006 [arXiv:1507.06308 [gr-qc]].



\bibitem{Oikonomou:2022bqb}
V.~K.~Oikonomou and I.~Giannakoudi,
Nucl. Phys. B \textbf{978} (2022), 115779
doi:10.1016/j.nuclphysb.2022.115779 [arXiv:2204.02454 [gr-qc]].

\bibitem{Lambiase:2022ucu}
G.~Lambiase, L.~Mastrototaro and L.~Visinelli,
JCAP \textbf{01} (2023), 011 doi:10.1088/1475-7516/2023/01/011
[arXiv:2207.08067 [hep-ph]].


\bibitem{reviews1}
 S.~Nojiri, S.~D.~Odintsov and V.~K.~Oikonomou,
  Phys.\ Rept.\  {\bf 692} (2017) 1
  [arXiv:1705.11098 [gr-qc]].

\bibitem{reviews2}


 S. Capozziello, M. De Laurentis,
   Phys.\ Rept.\  {\bf 509}, 167 (2011);\\




\bibitem{reviews3}
 V.~Faraoni and S.~Capozziello,
  Fundam.\ Theor.\ Phys.\  {\bf 170} (2010).


   \bibitem{reviews4}

S. Nojiri, S.D. Odintsov,
   Phys.\ Rept.\  {\bf 505}, 59 (2011);




\bibitem{reviews5}

G.~J.~Olmo,
  Int.\ J.\ Mod.\ Phys.\ D {\bf 20} (2011) 413
  [arXiv:1101.3864 [gr-qc]].






\bibitem{Nojiri:2003ft}
S.~Nojiri and S.~D.~Odintsov,
Phys.\ Rev.\ D {\bf 68} (2003) 123512
doi:10.1103/PhysRevD.68.123512 [hep-th/0307288].


\bibitem{Nojiri:2007as}
S.~Nojiri and S.~D.~Odintsov,
Phys.\ Lett.\ B {\bf 657} (2007) 238
doi:10.1016/j.physletb.2007.10.027 [arXiv:0707.1941 [hep-th]].

\bibitem{Nojiri:2007cq}
S.~Nojiri and S.~D.~Odintsov,
Phys.\ Rev.\ D {\bf 77} (2008) 026007
doi:10.1103/PhysRevD.77.026007 [arXiv:0710.1738 [hep-th]].

\bibitem{Cognola:2007zu}
G.~Cognola, E.~Elizalde, S.~Nojiri, S.~D.~Odintsov, L.~Sebastiani
and S.~Zerbini,
Phys.\ Rev.\ D {\bf 77} (2008) 046009
doi:10.1103/PhysRevD.77.046009 [arXiv:0712.4017 [hep-th]].

\bibitem{Nojiri:2006gh}
S.~Nojiri and S.~D.~Odintsov,
Phys.\ Rev.\ D {\bf 74} (2006) 086005
doi:10.1103/PhysRevD.74.086005 [hep-th/0608008].

\bibitem{Appleby:2007vb}
S.~A.~Appleby and R.~A.~Battye,
Phys.\ Lett.\ B {\bf 654} (2007) 7
doi:10.1016/j.physletb.2007.08.037 [arXiv:0705.3199 [astro-ph]].



\bibitem{Elizalde:2010ts}
E.~Elizalde, S.~Nojiri, S.~D.~Odintsov, L.~Sebastiani and
S.~Zerbini,
Phys.\ Rev.\ D {\bf 83} (2011) 086006
doi:10.1103/PhysRevD.83.086006 [arXiv:1012.2280 [hep-th]].



\bibitem{Sa:2020fvn}
P.~M.~S\'a,
Phys. Rev. D \textbf{102} (2020) no.10, 103519
doi:10.1103/PhysRevD.102.103519 [arXiv:2007.07109 [gr-qc]].




\bibitem{Ai:2020peo}
W.~Y.~Ai,
Commun. Theor. Phys. \textbf{72} (2020) no.9, 095402
doi:10.1088/1572-9494/aba242 [arXiv:2004.02858 [gr-qc]].



\bibitem{Hwang:2005hb}
J.~c.~Hwang and H.~Noh,
Phys. Rev. D \textbf{71} (2005), 063536
doi:10.1103/PhysRevD.71.063536 [arXiv:gr-qc/0412126 [gr-qc]].



\bibitem{Odintsov:2020sqy}
S.~D.~Odintsov, V.~K.~Oikonomou and F.~P.~Fronimos,
Nucl. Phys. B \textbf{958} (2020), 115135
doi:10.1016/j.nuclphysb.2020.115135 [arXiv:2003.13724 [gr-qc]].



\bibitem{Appleby:2009uf}
S.~A.~Appleby, R.~A.~Battye and A.~A.~Starobinsky,
JCAP \textbf{06} (2010), 005 doi:10.1088/1475-7516/2010/06/005
[arXiv:0909.1737 [astro-ph.CO]].






\bibitem{Oikonomou:2020oil}
V.~K.~Oikonomou and F.~P.~Fronimos,
EPL \textbf{131} (2020) no.3, 30001
doi:10.1209/0295-5075/131/30001 [arXiv:2007.11915 [gr-qc]].



\bibitem{Oikonomou:2020krq}
V.~K.~Oikonomou,
EPL \textbf{130} (2020) no.1, 10006
doi:10.1209/0295-5075/130/10006 [arXiv:2004.10778 [gr-qc]].



\bibitem{Planck:2018vyg}
N.~Aghanim \textit{et al.} [Planck],
Astron. Astrophys. \textbf{641} (2020), A6 [erratum: Astron.
Astrophys. \textbf{652} (2021), C4]
doi:10.1051/0004-6361/201833910 [arXiv:1807.06209 [astro-ph.CO]].



\bibitem{Maldacena:2002vr}
J.~M.~Maldacena,
JHEP \textbf{05} (2003), 013 doi:10.1088/1126-6708/2003/05/013
[arXiv:astro-ph/0210603 [astro-ph]].


\bibitem{Caldwell:2003vq}
R.~R.~Caldwell, M.~Kamionkowski and N.~N.~Weinberg,
Phys. Rev. Lett. \textbf{91} (2003), 071301
doi:10.1103/PhysRevLett.91.071301 [arXiv:astro-ph/0302506
[astro-ph]].



\bibitem{Kinney:2018nny}
W.~H.~Kinney, S.~Vagnozzi and L.~Visinelli,
Class. Quant. Grav. \textbf{36} (2019) no.11, 117001
doi:10.1088/1361-6382/ab1d87 [arXiv:1808.06424 [astro-ph.CO]].



\bibitem{Kehagias:2018uem}
A.~Kehagias and A.~Riotto,
Fortsch. Phys. \textbf{66} (2018) no.10, 1800052
doi:10.1002/prop.201800052 [arXiv:1807.05445 [hep-th]].



\bibitem{Achucarro:2018vey}
A.~Ach\'ucarro and G.~A.~Palma,
JCAP \textbf{02} (2019), 041 doi:10.1088/1475-7516/2019/02/041
[arXiv:1807.04390 [hep-th]].









\end{thebibliography}
\end{document}